\documentclass[12pt]{article} 
\usepackage[utf8x]{inputenc} 
\usepackage[T1]{fontenc} 
\usepackage{amsmath,amssymb,bbm,commath,xfrac} 
\usepackage{sectsty} 
\usepackage{graphicx} 
\usepackage{color,xparse} 
\usepackage[unicode]{hyperref} 
\hypersetup{bookmarksnumbered=true, bookmarksopen=true, bookmarksopenlevel=2, breaklinks=true, citecolor=blue, colorlinks=true, linkcolor=red, linktoc=page, pdfborder={0 0 0}, pdfkeywords={3D N≥2, }{Chern-Simons-Matter, }{Quiver Gauge Theories}, pdfstartview=FitH, pdfauthor={DJain}, pdfsubject={Computation of Free Energy of 3D Quiver Gauge Theories}, pdftitle={Deconstructing ζ-Deformed D-quivers}, plainpages=false, unicode=true, urlcolor=cyan}
\usepackage{cite} 

\usepackage{geometry} 
\usepackage{fancyhdr} 
\usepackage{parskip} 
\usepackage[titles]{tocloft} 

\DeclareUnicodeCharacter{"0393}{\Gamma}
\DeclareUnicodeCharacter{"0394}{\Delta}
\DeclareUnicodeCharacter{"0398}{\Theta}
\DeclareUnicodeCharacter{"039B}{\Lambda}
\DeclareUnicodeCharacter{"039E}{\Xi}
\DeclareUnicodeCharacter{"03A0}{\Pi}
\DeclareUnicodeCharacter{"03A3}{\Sigma}
\DeclareUnicodeCharacter{"03A5}{\Upsilon}
\DeclareUnicodeCharacter{"03A6}{\Phi}
\DeclareUnicodeCharacter{"03A8}{\Psi}
\DeclareUnicodeCharacter{"03A9}{\Omega}
\DeclareUnicodeCharacter{"03B1}{\alpha}
\DeclareUnicodeCharacter{"03B2}{\beta}
\DeclareUnicodeCharacter{"03B3}{\gamma}
\DeclareUnicodeCharacter{"03B4}{\delta}
\DeclareUnicodeCharacter{"03B5}{\epsilon}
\DeclareUnicodeCharacter{"03B6}{\zeta}
\DeclareUnicodeCharacter{"03B7}{\eta}
\DeclareUnicodeCharacter{"03B8}{\theta}
\DeclareUnicodeCharacter{"03D1}{\vartheta}
\DeclareUnicodeCharacter{"03B9}{\iota}
\DeclareUnicodeCharacter{"03BA}{\kappa}
\DeclareUnicodeCharacter{"03BB}{\lambda}
\DeclareUnicodeCharacter{"03BC}{\mu}
\DeclareUnicodeCharacter{"03BD}{\nu}
\DeclareUnicodeCharacter{"03BE}{\xi}
\DeclareUnicodeCharacter{"03C0}{\pi}
\DeclareUnicodeCharacter{"03C1}{\rho}
\DeclareUnicodeCharacter{"03C3}{\sigma} 
\newcommand\s {\sigma}
\DeclareUnicodeCharacter{"03C4}{\tau}
\DeclareUnicodeCharacter{"03C5}{\upsilon}
\DeclareUnicodeCharacter{"03C6}{\phi}
\DeclareUnicodeCharacter{"03D5}{\varphi}
\DeclareUnicodeCharacter{"03C7}{\chi}
\DeclareUnicodeCharacter{"03C8}{\psi}
\DeclareUnicodeCharacter{"03C9}{\omega}
\DeclareUnicodeCharacter{"21D0}{\Leftarrow}
\DeclareUnicodeCharacter{"0212B}{\AA}
\DeclareUnicodeCharacter{"00B7}{\cdot}
\DeclareUnicodeCharacter{"266A}{\eighthnote}
\DeclareUnicodeCharacter{"266B}{\twonotes}
\PrerenderUnicode{ä}\PrerenderUnicode{×}

\definecolor{title}{rgb}{0.1,0.5,0.9}
\definecolor{abst}{rgb}{0.366,0.366,0.266}
\definecolor{sect}{rgb}{0.1,0.3,0.8}
\definecolor{ssect}{rgb}{0.1,0.4,0.7}
\definecolor{sssect}{rgb}{0.3,0.3,0.3}
\definecolor{appsect}{rgb}{0.1,0.1,0.5}
\definecolor{ref}{rgb}{0.0,0.0,1.0}
\newcommand{\Title}[1] {\title{\color{title}\Huge #1}}
\newcommand{\TPheader}[3] {\thispagestyle{fancy}\pagenumbering{alph}\lhead{#1}\chead{#2}\rhead{#3}\cfoot{}}
\newcommand{\makepage}[1] {\newpage\pagenumbering{#1}}

\newcommand{\Abstract}[1] {\begin{abstract}\normalsize #1 \end{abstract}}

\sectionfont{\color{sect}\Large\bf}
\subsectionfont{\color{ssect}\large\bf}
\subsubsectionfont{\color{sssect}\normalsize\bf}
\renewcommand{\appendix}{\setcounter{section}{0}\sectionfont{\color{appsect}\Large\bf}\renewcommand{\thesection}{\Alph{section}}\renewcommand*{\theHsection}{app.\the\value{section}}} 

\newcommand\references[1] {\begin{thebibliography}{[99]}\addcontentsline{toc}{section}{References} #1 \end{thebibliography}}
\NewDocumentCommand\arXivid{m+g}{\IfNoValueTF{#2}{\href{http://arxiv.org/abs/#1}{\tt arXiv:#1}}{\href{http://arxiv.org/abs/#1}{\tt arXiv:#1{\small [#2]}}}}
\newcommand\cmp[3] {{\it Commun.\ Math.\ Phys.\ } \href{http://inspirehep.net/search?ln=en&ln=en&p=find+j+"Commun.Math.Phys.,#1,#3"&of=hb&action_search=Search&sf=&so=d&rm=&rg=25&sc=0}{{\bf #1} (#2) #3}} 

\newcommand\jhep[3]{{\it JHEP\ } \href{http://inspirehep.net/search?ln=en&ln=en&p=find+j+"JHEP,#1,#3"&of=hb&action_search=Search&sf=&so=d&rm=&rg=25&sc=0}{{\bf #1} (#2) #3}}
\newcommand\npb[3] {{\it Nucl.\ Phys.\ }{\bf B #1} (#2) #3}
\newcommand\pr[4] {{\it Phys.\ Rev.\ }{\bf #1 #2} (#3) #4} 

\newcommand\eqs[1] {\begin{align}#1\end{align}}
\newcommand\eqsn[1] {\begin{align*}#1\end{align*}}

\newcommand\equ[1] {\begin{equation}#1\end{equation}}
\newcommand\equn[1] {\begin{equation*}#1\end{equation*}}
\newcommand\fig[2] {\begin{figure}[#1]\centering #2\end{figure}}

\renewcommand\exp[1] {e^{#1}}

\newcommand\half {\tfrac{1}{2}}
\renewcommand\( {\left(}
\renewcommand\) {\right)}
\newcommand\wh {\widehat}

\renewcommand\C {{\cal C}}

\newcommand\N {{\cal N}}

\renewcommand\P {{\cal P}}

\newcommand\bC {{\mathbb C}}
\newcommand\bD {{\mathbb D}}
\newcommand\bR {{\mathbb R}}
\newcommand\bZ {{\mathbb Z}}

\newcommand\bs[1] {\boldsymbol{#1}} 
\newcommand\ie {\textit{i.e.}}

\newcommand\nn {\nonumber\\}

\numberwithin{equation}{section} 
\begin{document}
\Title{\bf Deconstructing $\bs{ζ}$-Deformed $\bs{\wh{D}}$-quivers}
\author{Dharmesh Jain\footnote{\href{mailto:djain@phys.ntu.edu.tw}{djain@phys.ntu.edu.tw}}\bigskip\\
\emph{Department of Physics and Center for Theoretical Sciences}\\
\emph{National Taiwan University, Taipei 10617, Taiwan}
}
\date{} 

\maketitle
\TPheader{\today}{}{} 

\Abstract{We extend our previous analysis of $d=3, \N=3$ supersymmetric Chern-Simons-matter theories of affine quiver types by including the Yang-Mills action and non-vanishing (complex) FI parameters (which break susy to $\N=2$). We find that they can be interpreted as giving rise to non-canonical R-charges for the bifundamental fields. This leads to some straightforward generalizations of the `canonical' volume/free energy (as in AdS/CFT) formulas and the cone construction for those volume formulas.

$\hphantom{sentence}$\\
}

\tableofcontents

\makepage{arabic}
\section{Introduction}
Year 2015 saw publication of two papers \cite{Moriyama,Drukker} dealing with non-trivial extensions of ABJM theories, namely $\wh{D}$-quivers. They calculated the free energy explicitly using similar approaches (Fermi gas formalism\cite{FGF1,FGF2}) and verified the result (in large $N$ limit) we `conjectured' in \cite{CHJ}. The free energy of such field theories, defined by logarithm of the sphere partition function, is related to the gravitational free energy via AdS/CFT as follows \cite{DMP,HKPT}
\equ{F=-\ln Z_{S^3}=N^{\sfrac{3}{2}}\sqrt{\frac{2 \pi^6}{27\,\text{Vol}(Y^7)}}+o(N^{\sfrac{3}{2}}),
\label{F Vol}
}
where $\text{Vol}(Y^7)$ is the volume of a 7-dimensional (tri-)Sasaki-Einstein manifold. It is believed that the M-theoretical description of these field theories arises as the near-horizon limit of a stack of $N$ M2-branes located at the tip of a 8-dimensional Calabi-Yau cone with $Y^7$ as its base.

There have also been a lot of work dealing with trivial deformations and hence computations of partition functions and/or study of phases of ABJM (and related) theories. For an incomprehensive list of papers from last year, see \cite{Anderson:2015ioa,Hatsuda:2015gca,Russo:2015exa,Nosaka:2015bhf,Nosaka:2015iiw}. So we decided to merge these recent developments and considered calculating the free energy of $\wh{D}$-quivers deformed by FI parameters. We find that by using complex FI parameters, the free energy expressions can be interpreted as coming from a quiver theory with vanishing FI paramters but with non-canonical R-charges assigned to the bifundamental matter fields (along the lines discussed in \cite{Nosaka:2015bhf} for ABJM theory).

In the next section, we give a brief review of the field theory setup to be discussed here. Then in section \ref{solution}, we will give the algorithm to solve (and solution of) the matrix models for $\wh{D}$-quivers, which allows us to write down the general volume formula with non-zero FI parameters. We will also discuss how to get the (non-canonical) R-charges for the bifundamental matter fields out of these expressions. Finally, before we conclude in section \ref{discussion}, we do the same analysis for $\wh{A}$-quivers in appendix \ref{Asol Appendix}. Appendix \ref{A and D Appendix} contains our conventions for roots of affine $A$ and $D$ Lie algebras, which are used to relabel CS levels for the corresponding quiver diagrams.

\section{Review}\label{review}
We will consider quiver Chern-Simons-Yang-Mills gauge theories involving products of unitary groups, \ie, $G=\otimes_a U(N_a)$, coupled to bifundamental hypermultiplets and some fundamentals. According to \cite{KWY}, the partition function of these theories on $S^3$ is localized on configurations where the auxiliary scalar fields $\sigma_a$ in the $\mathcal N=2$ vector multiplets are  constant $N \times N$ matrices. Thus, evaluating the free energy amounts to solving a matrix model. Also, the free energy depends non-trivially on two sets of parameters in the theory: CS levels $k_a$ and FI parameters $ζ_a$ associated with a node `$a$'. (The labelling of nodes for $\wh{D}$-quivers is shown in Figure \ref{fig:Dn}.)
\fig{h!}{\includegraphics[width=3in]{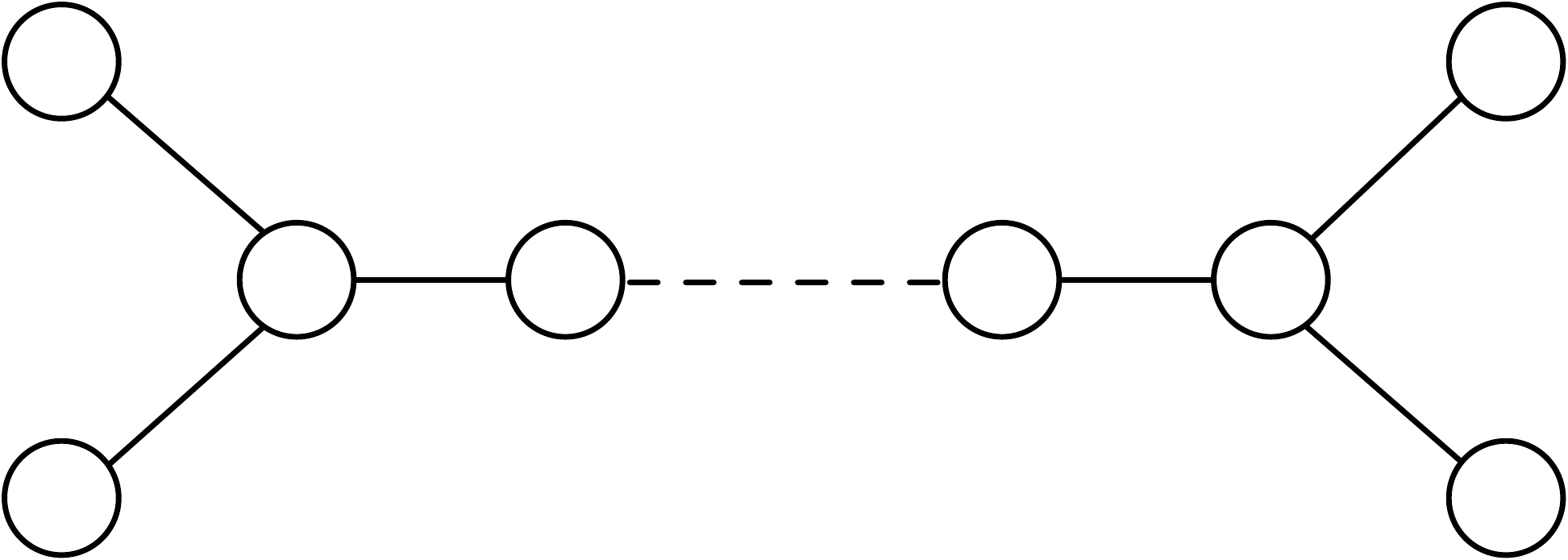}
\put(1,4){$k_3$}\put(1,65){$k_4$}\put(-228,65){$k_1$}\put(-228,4){$k_2$}\put(-180,50){$k_5$}\put(-142,50){$k_6$}\put(-58,50){$k_{n+1}$}
\caption{$\widehat D_n$ quiver diagram with the CS levels marked as per our convention.}
\label{fig:Dn}
}

A few comments on symmetries: Base supersymmetry of these models is $\N=2$, which may be enhanced to $\N=3,⋯,8$ under suitable conditions like $ζ→0$ and/or particular choices of CS levels and/or particular choices of gauge groups. These models are not (super)conformal as we will see later that the nonvanishing FI parameters can be interpreted as non-canonical R-charges for the bifundamental fields. There is also a mirror duality (for $\N=4$) in IR which relates $ζ$'s and fundamental hypermultiplet masses $m$'s \cite{HW,KapD,ADey} implying that the two mirror theories are related by an exchange of their Coulomb and Higgs branch. Even if the theories we consider are not at their superconformal fixed point(s), the free energy formula \eqref{F Vol} continues to hold as discussed in \cite{MaSp1,MaSp2,JKPS}. In addition, given such a free energy formula, one can consider `extremizing' it with respect to the R-charges (subject to some constraints like their sum takes a particular value and so on) and obtain the free energy at an IR fixed point along some (un)known RG flow (see \cite{JKPS} for many examples in ABJM-type theories). We will not discuss these issues any further and point the reader to the literature (some of which we already mentioned) and/or leave it as a homework exercise.

\subsection*{Matrix Model}
We denote the eigenvalues of $\sigma_a$ in each vector multiplet by $\lambda_{a,i}$, $i=1,...,N_a$. The sphere partition function is then given by
\equ{Z_{S^3} = \int \Bigg( \prod_{a, i} d \lambda_{a,i} \Bigg) L_v(\{\lambda_{a,i}\}) L_m(\{\lambda_{a,i}\}) = \int \Bigg( \prod_{a, i} d \lambda_{a,i}\Bigg) \exp{-F (\{ \lambda_{a,i} \})},
}
where the contribution from vector multiplets is
\equ{L_v = \prod_{a=1}^d \frac{1}{N_a!} \Bigg( \prod_{i > j} 2 \sinh[ \pi (\lambda_{a,i} - \lambda_{a,j})] \Bigg)^2 \exp{2π i ∑_{a,j}ζ_a λ_{a,j} + \pi i \sum_{a,j} k_a \lambda_{a,j}^2},
}
and from matter multiplets is
\equ{L_m = \prod_{(a,b) \in E} \prod_{i,j} \frac{1}{2 \cosh [ \pi (\lambda_{a,i} - \lambda_{b,j})]} \prod_c \Bigg(\prod_i \frac{1}{2 \cosh[\pi \lambda_{c,i}]}\Bigg)^{n^f_c}.
} 
The first product in $L_m$ is due to the bifundamental fields while the second one is due to the fundamental flavour fields, where $n^f_c$ is the number of pairs of these flavour fields at the node labelled by index $c$.

\subsection*{Large $\boldsymbol{N}$ Limit and $\boldsymbol{\widehat{ADE}}$ Classification}
Following \cite{HKPT,Gulotta2}, we assume that the eigenvalue distribution becomes dense in the large $N$ limit, \ie, $\lambda_{a,i} \rightarrow \lambda_a(x)$ with a certain density $\rho(x)$. In this limit  the free energy becomes a 1-dimensional integral which we evaluate by saddle point approximation. We also assume that the eigenvalue distribution for a node with $N_a=n_a N$ is given by a collection of $n_a$ curves in the complex plane labelled by $\lambda_{a,I}(x)$ with $I=1, ...,n_a$ and write the ansatz 
\equ{\lambda_{a,I}(x)=N^\alpha x +i\, y_{a,I}(x).
\label{Ansatz lambda}
}
The eigenvalue density at each node is assumed to be given by $\rho(x)$, which satisfies
\equ{\textstyle \int dx \rho(x)=1.
\label{Normalization rho}
}
This normalization condition will be imposed through a Lagrange multiplier $\mu$ below. As explained in \cite{Gulotta2}, the leading order in $N$ in the saddle point equation is proportional to the combination $2 n_a -\sum_{b| (a,b)\in E} n_b$. The requirement that this term vanishes is equivalent to the quiver being in correspondence with the simply laced extended Dynkin diagrams, leading to the well-known $\wh{ADE}$ classification.

To leading order in $N$, the free energy contains a tree-level contribution and a 1-loop contribution. Assuming $\sum_a n_a k_a=0$ and the requirement that these two contributions are balanced leads to $\alpha=\half$, which is ultimately responsible for the $N^{\sfrac{3}{2}}$ scaling of the free energy. Finally, the free energy to be extremized reads
\eqs{F &= N^{\sfrac{3}{2}}\int \rho(x) \Biggl[π n_F |x|+ 2 π x\, t_ζ + 2 π x \sum_a \sum_{I=1}^{n_a} k_a y_{a,I}(x) \nn
&\qquad + \frac{\rho(x)}{4\pi} \Biggl(\sum_{a=1}^d \sum_{I=1}^{n_a} \sum_{J=1}^{n_a} \arg \(e^{2 π i (y_{a,I} - y_{a,J} - \sfrac{1}{2})} \)^2 -\sum_{(a,b) \in E} \sum_{I=1}^{n_a} \sum_{J=1}^{n_b} \arg \( e^{2 π i (y_{a,I} - y_{b,J})} \)^2\Biggr) \Biggr] dx \nn
&\qquad -2 π μ N^{\sfrac{3}{2}} \(\int \rho(x)\, dx -1\)\!,
\label{F}
}
where $n_F \equiv \sum_a n_a n_a^f$ and $t_ζ=∑_a n_at_a$ such that $t_a≡\textit{Im}(ζ_a)$. We note that the real part of $ζ_a$ does not appear in $F$ at the leading order in $N$ (a fact verifiable via `numerical experiments'). Also, notice that the integrand for free energy is no longer symmetric along the `$x$-axis' due to the appearance of $t_ζ$ term. This is a non-trivial statement whose first consequence is that the functions $ρ(x)$ and $y(x)$ are not symmetric anymore under $x→-x$. Second serious consequence is that our algorithm of \cite{CHJ} for solving such matrix models is no longer directly applicable and we have to take care of the `negative' $x$-axis too. However, evaluating the free energy on-shell still gives the same simple relation
\equ{F=\frac{4 \pi N^{\sfrac{3}{2}}}{3} \mu,
\label{Fproptomu}
}
as proven earlier in \cite{Gulotta1}. Thus, $F$ is determined by $\mu$, which in turn is determined as a function of the CS levels and FI parameters from the normalization condition (\ref{Normalization rho}). Note that from (\ref{F Vol}) and (\ref{Fproptomu}), it follows that
\equ{\frac{\text{Vol}(Y^7)}{\text{Vol}(S^7)}=\frac{1}{8\mu^2}·
\label{relation vol and mu}
}

It is convenient to relate the CS level $k_{(a)}$ at each node  to a root $\alpha_a$, by  introducing a ($n$-dimensional) vector $p$ and writing $k_{(a)}=\alpha_a \cdot p\,$. This way, the condition $\sum_a n_a k_a=0$ is satisfied automatically. See Appendix~\ref{A and D Appendix} for an explicit basis.

\section{Solution}\label{solution}
This is the main section where we discuss the {\bf algorithm} to solve the matrix model for deformed $\wh{D}$-quivers and present the general {\bf volume} formula. We also present the associated {\bf cone} construction and relate the parameter $t_ζ$ to non-canonical {\bf R-charges} of the bifundamental fields. Let us proceed in this precise order.

\subsection*{Algorithm}
Following our algorithm from \cite{CHJ}:
\emph{Extremizing (\ref{F}) (with respect to $y_{a,I}$ and $\rho$) requires an assumption on the branch of the arg functions. We will always take the principle value and therefore we assume that
\eqs{|y_{a,I} - y_{a,J}| < 1\,; \qquad | y_{a,I} - y_{b,J}| < \tfrac{1}{2}\,, \quad \text{if} \,\, (a,b) \in E.
\label{inequalities}
}
Based on numerical results \cite{HKPT,Gulotta2}, we assume that the $n_a$ curves for a given node initially coincide, \ie, $|y_{a,I}-y_{a,J}|=0$. Extremizing $F$ under these assumptions, one finds that the solution is consistent only in a bounded region away from the origin. This is because as $|x|$ increases, the differences $|y_{a,I} - y_{b,J}|$ monotonically increase (or decrease), saturating  one (or more) of the inequalities assumed in (\ref{inequalities}) at some point.}

This is the place where our previous algorithm fails because saturation on positive $x$-axis no longer implies saturation on negative $x$-axis due to the notorious asymmetry of the integrand in $F$. However, all is not lost and after considering a few cases, a consistent choice can be made for saturation points $x_*^-$ on negative $x$-axis from the knowledge of $x_*^+$ only. Here's the rule:
\equ{\text{If } x_*^+=μ\,X\big(n_F,t_ζ;\{k_a\}\big) \text{ then } x_*^-=-μ\,X\big(n_F,-t_ζ;\{k_a\}\big).
}
This rule may seem to be picked out of thin air but its origin will become clear in a while. After this, our old algorithm can proceed without (m)any problems.

\emph{The relation among the CS levels determines the sequence in which these inequalities saturate. This saturation will be maintained beyond this point, requiring the eigenvalue distribution involved either to bifurcate or develop a kink. After a saturation occurs, the total number of independent variables is reduced by one. Thus, at this point, we remove one variable from the Lagrangian, revise the inequalities and solve the equations of motion again until a new saturation is encountered. This process is iterated until all $y_a$'s are related, determining a maximum of $(\sum_a n_a-1)$ regions or until the eigenvalue distribution terminates, \ie, $\rho(x)=0$.  Once the eigenvalue density $\rho(x)$ is determined in all regions, the value of $\mu$ (and therefore $F$) is found from the normalization condition (\ref{Normalization rho}).}

We apply this modified algorithm first for $\wh{A}$ series as a sanity/consistency check. A sample plot of eigenvalue density for $\wh{A}_2$ is shown in Figure \ref{fig:A2}.
\fig{h!}{\includegraphics[scale=0.5]{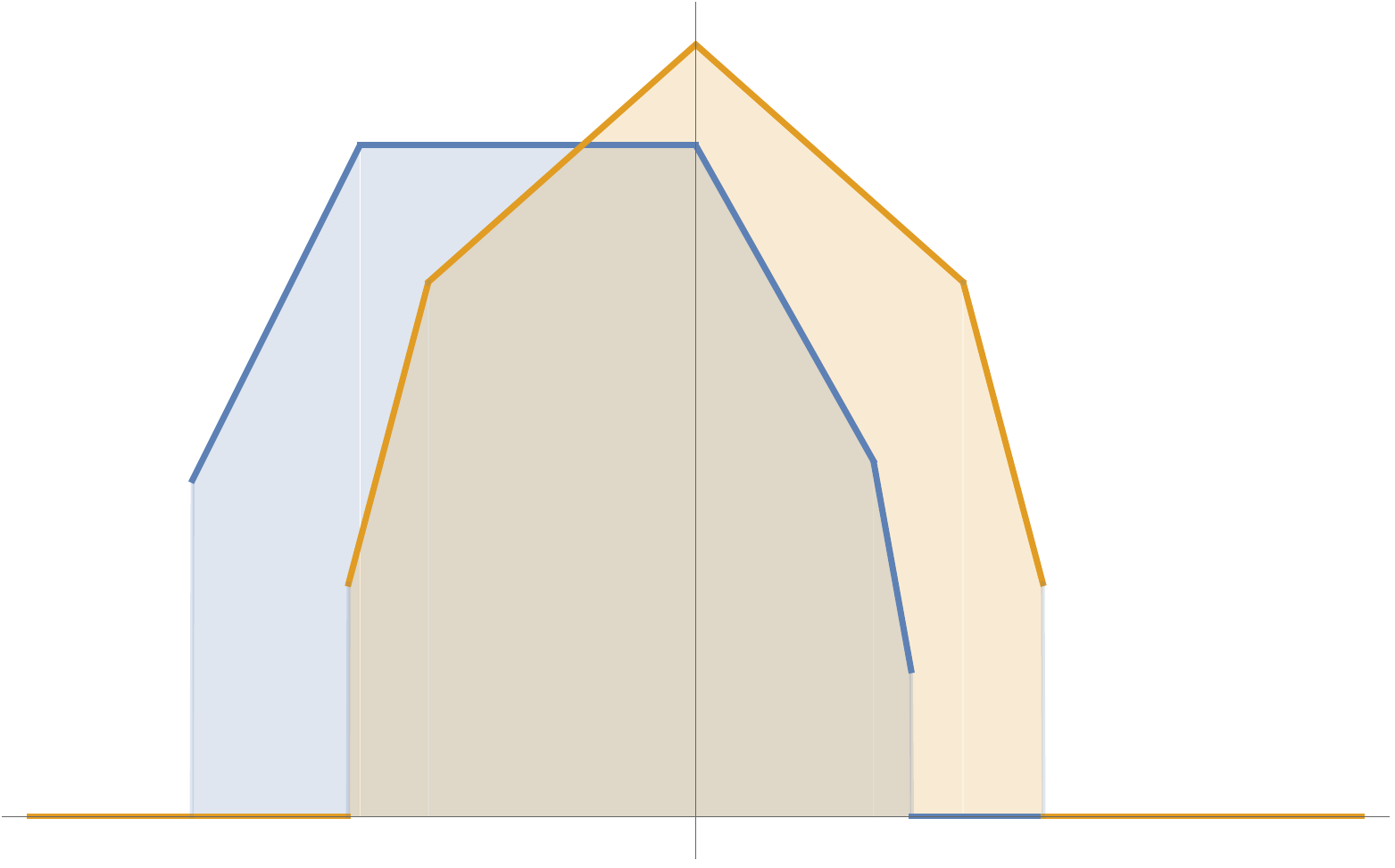}
\caption{Eigenvalue density $ρ(x)$ for $\wh{A}_2$ with $t_ζ=0$ (orange) and $t_ζ≠0$ (blue).}
\label{fig:A2}
}
But the great insight about $t_ζ$ is obtained by considering $\wh{A}_1$ quiver following \cite{JKPS,Nosaka:2015bhf}. It is revealed that this parameter can be related to non-canonical R-charges of the bifundamental fields. The specific relation being
\equ{Δ_A=\half -\tfrac{t_ζ}{k}\,;\qquad Δ_B=\half +\tfrac{t_ζ}{k},
}
where $A$ denotes a bifundamental field in representation $(\bs{N},\bs{\bar{N}})$ and $B$ denotes $(\bs{\bar{N}},\bs{N})$. Of course, when $ζ$'s vanish, $Δ$'s reduce to their canonical value of $\half$. We relegate the rest of the discussion for $\wh{A}$-quivers to the Appendix \ref{Asol Appendix} because generalizing solution for $\wh{D}$-quivers is much more straightforward. We will assume that $Δ$'s and $t_ζ$ can be similarly related in the case of $\wh{D}$-quivers too and eventually conjecture the leading order of $F$ in terms of these R-charges.

The simplest example from $\wh{D}$ series is the $\wh{D}_4$ quiver whose essence is captured by the Figures \ref{fig:D4rho} and \ref{fig:D4ys}.
\fig{h!}{\includegraphics[scale=0.45]{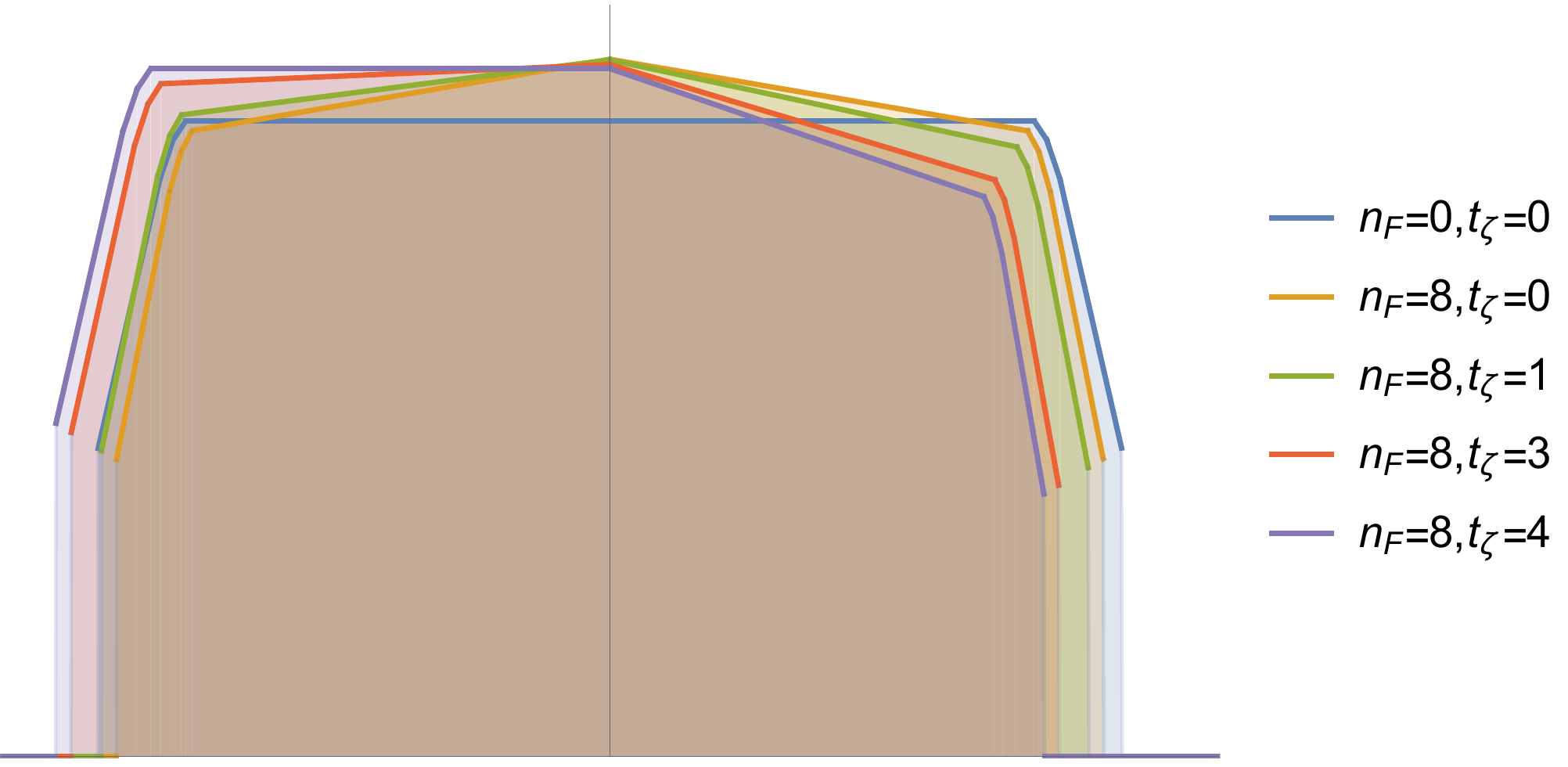}
\caption{Eigenvalue density $ρ(x)$ for $\wh{D}_4$ with $n_F=\{0,\,8\}$ and $t_ζ\in\{0,4\}$.}
\label{fig:D4rho}
}
\fig{h!}{\includegraphics[scale=0.4]{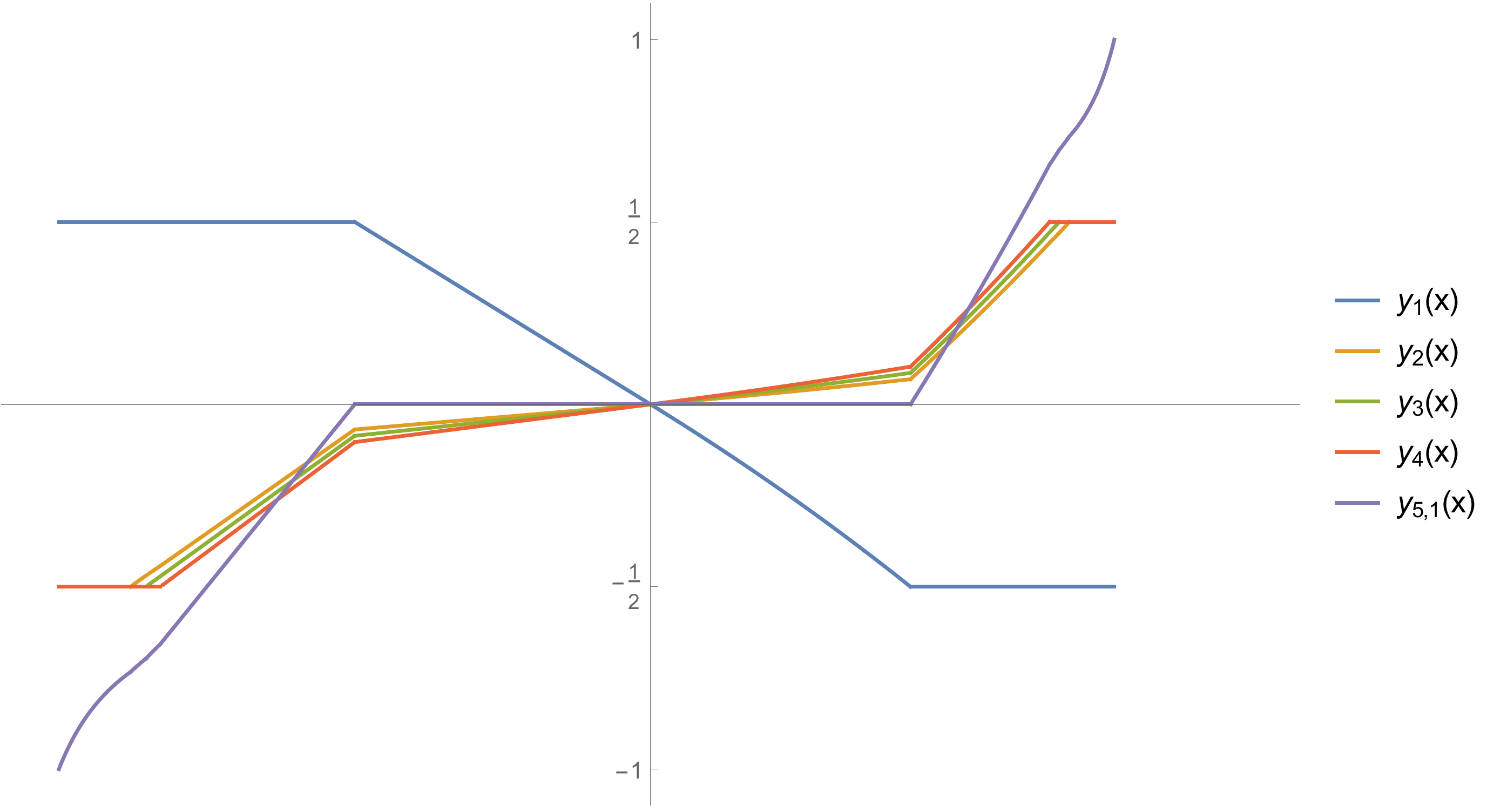}
\caption{(Asymmetric) $y_{a,I}(x)$'s for $\wh{D}_4$ with $n_F=8$ and $t_ζ=4$. (We have set $y_{5,2}(x)=0$.)}
\label{fig:D4ys}
}

\noindent  As discussed, $ρ(x)$ and $y(x)$'s are clearly no longer symmetric across the $x$-axis. Now we turn to the simpler task of integrating the density to get the volume formula.

\subsection*{Volume}
We think that giving some test cases is always a good idea before unleashing the grandeur of a general formula so here are two expressions for volumes of $\wh{D}_4$ \eqref{Sol D4} and $\wh{D}_5$ \eqref{Sol D5}:
\eqs{\frac{1}{8\mu^2} &=\frac{1}{16}\left[-\frac{1}{n_F+2t_ζ+4 p_1}+\frac{4(n_F+2t_ζ+2 p_1+3 p_2-p_3)}{\big(n_F+2t_ζ+2(p_1+p_2)\big)^2}-\frac{1}{n_F+2t_ζ+2(p_1+p_2+p_3-p_4)}\right. \nn
&\quad\left.-\frac{1}{n_F+2t_ζ+2(p_1+p_2+p_3+p_4)}+\(t_ζ↔-t_ζ\)\right]\!· \label{Sol D4} \\
\frac{1}{8\mu^2} &=\frac{1}{48}\left[-\frac{1}{n_F+2t_ζ+6 p_1} -\frac{3}{n_F+2(t_ζ+p_1 + 2 p_2)} +\frac{12(n_F+2t_ζ+2 p_1 + 2 p_2 + 3 p_3 - p_4)}{\big(n_F+2(t_ζ+p_1 + p_2 + p_3)\big)^2}\right. \nn
&\quad-\left.\frac{3}{n_F+2(t_ζ+p_1 + p_2 + p_3 + p_4 - p_5)} -\frac{3}{n_F+2(t_ζ+p_1 + p_2 + p_3 + p_4 + p_5)}+(⋯)\right]\!· \label{Sol D5}
}
We have used the inequalities: $k_{n+1}≥k_n≥⋯≥k_2≥0$ or equivalently $p_1≥p_2≥⋯≥p_n$ in our explicit calculations.

We recall from \cite{CHJ} that the above volume formula can be generalized in terms of $(n+3)$ 2-dimensional vectors $β$'s. Here, we will need $β_a= (1, p_a)$ together with $\beta_0=(0,1)$, (modify $β_F$ to) $β^±_{\sfrac{1}{2}}=\(0,\frac{n_F}{2}±t_ζ\)$, and $\beta_{n+1}=(1,0)$. Defining the wedge product $(a,b) \wedge (c,d) = (a d- b c)$, we can introduce modified $\bar \sigma_a$'s in terms of $\gamma_{a,b}\equiv \left|  \beta_a \wedge \beta_b \right|$ as follows
\equ{\bar \sigma_a^{±}  = \sum_{b=\sfrac{1}{2},1}^n \( \gamma_{a,b}+ \gamma_{a,-b} \) - 4 \gamma_{a,n+1}\,; \quad a=0,\sfrac{1}{2},1,⋯,n+1,
\label{all sigmas}
}
where $\beta_{-a}$ denotes negating the second component of $β_a$. Obviously, $±$ on $\bar{\s}_a$ refers to the presence (if any) of $β^±_{\sfrac{1}{2}}$ on the right-hand side. This finally leads to the compact volume formula for $\wh{D}$-quivers
\equ{\frac{\text{Vol}(Y^7)}{\text{Vol}(S^7)}=\frac{1}{4}\(\sum_{a=0,1}^{n} \frac{\gamma_{a,a+1}}{\bar \sigma_a^-\, \bar \sigma_{a+1}^-} +\sum_{a=0,1}^{n} \frac{\gamma_{a,a+1}}{\bar \sigma_a^+\, \bar \sigma_{a+1}^+}\)\!·
\label{Dn sigmas}
}
Note that this sum does not contain $a=\sfrac{1}{2}$ because of its redundancy as $γ_{0,\sfrac{1}{2}}=0$ and the fact that terms with $γ_{\sfrac{1}{2},1}$ give the same contributions as $γ_{0,1}$. The two sums above are a consequence of $ρ(x)$ not being reflection-symmetric across $x$-axis. Of course when all $ζ_a=0$, $\bar{\s}_a^±$ reduce to $\bar{\s}_a$ of \cite{CHJ} and the overall factor adds up to a $\half$.

Setting all $ζ$'s and $k$'s to zero above, we simply get (from $a=0$ only) for the free energy
\equ{F=\frac{4πN^{\sfrac{3}{2}}}{3}\sqrt{\frac{4(n-2)n_F}{8}}=\frac{4πN^{\sfrac{3}{2}}}{3}\sqrt{(n-2)\frac{n_F}{2}}\,,
}
which exactly matches the large $N$ limit of the `Airy function' derived using Fermi gas formalism in \cite{Drukker}:
\equ{F=\frac{4πN^{\sfrac{3}{2}}}{3}\sqrt{Lν},
}
when we relate $n=L+2$ and $ν=\frac{n_F}{2}$. As mentioned there, these theories are dual to $2N$ $M2$-branes at an orbifold singularity in $\bC^2/\bZ_{n_F}×\bC^2/\bD_L$ such that the backreacted geometry in large $N$ limit is $AdS_4×S^7/(\bZ_{n_F}×\bD_L)$. So
\[\frac{\text{Vol}(Y^7)}{\text{Vol}(S^7)}=\frac{1}{8μ^2}=\frac{1}{8Lν}=\frac{1}{(n_F)(4L)}\,,\]
which matches with the above $(A,D)$-orbifolding of $S^7$.\footnote{We get a similar volume formula for $\wh{A}_{m-1}$ quivers (see Appendix \ref{Asol Appendix}) whose associated $Y^7≡S^7/(\bZ_{n_F}×\bZ_m)$ \cite{Zaffaroni,ADey} gives
\[\frac{\text{Vol}(Y^7)}{\text{Vol}(S^7)}=\frac{1}{(n_F)(m)}=\frac{1}{8\(\frac{mν}{4}\)}=\frac{1}{8μ^2}·\]
}

\subsection*{Cone}
In our previous paper \cite{CHJ}, we also gave a (convoluted) cone construction for the volume formula, such that each term ($\frac{γ}{\bar{\s}\bar{\s}}$) corresponded to the area of a triangle used to triangulate the cone. Generalizing that construction to the above $ζ$-deformed volume is straightforward but we can do better here by writing down one simple equation describing the complete polygon (not just the cone). This is done by generalizing the `cone' construction pioneered in the Fermi gas formalism for `pure' $\wh{D}$-quivers \cite{Moriyama}:
\equ{\begin{gathered}
\P=\left\{(x,y) ∈ \bR^2\left|{\textstyle\sum_{j=1}^{n}}\(|y+p_j x| +|y -p_j x|\) +2\left|\tfrac{n_F}{2}x+t_ζ |x|\right| - 4|y| ≤ \tfrac{1}{\sqrt{2}}\right.\right\}; \\
\text{Area}(\P)=2\,\frac{\text{Vol}(Y^7)}{\text{Vol}(S^7)}·
\end{gathered}
\label{Dcone}
}
Figure \ref{fig:D4cone} shows some polygons (and a cone) for $\wh{D}_4$ quiver.
\fig{h!}{\includegraphics[scale=0.4]{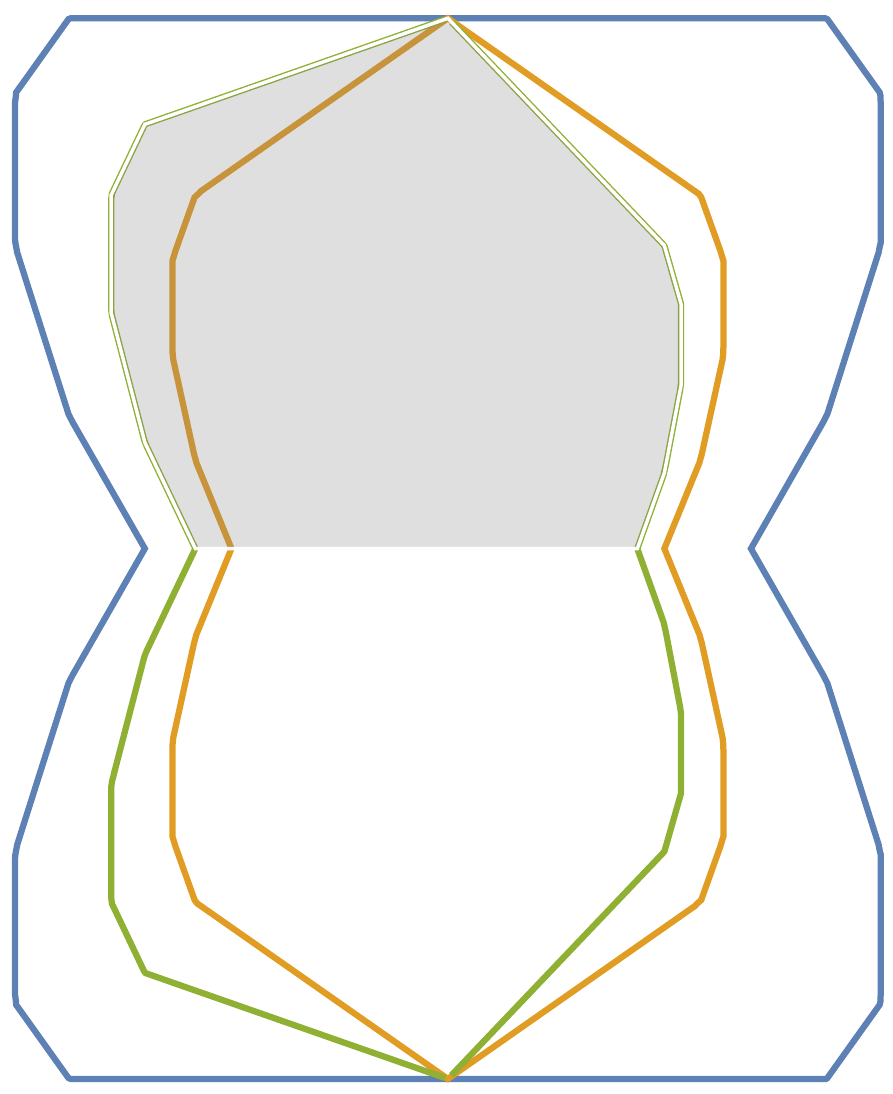}
\caption{[Blue] $\P$ for pure $\wh{D}_4$; [Orange] $\P$ (still symmetric) for flavoured $\wh{D}_4$; [Green] $\P$ (notice the asymmetry) for flavoured$+$deformed $\wh{D}_4$; [Grey] $\C$ using \cite{CHJ}'s original construction such that $\text{Area}(\C)$ gives the exact volume formula for $\wh{D}_4$.}
\label{fig:D4cone}
}

It should not be surprising if folding/unfolding procedure works as before (but we won't discuss it here). What we will emphasize here is that the `$F$-theorem' clearly holds for these $ζ$-deformed quivers. This is because the cone expands to cover more area as $n_F→0$ and/or $ζ→0$, which leads to a decrease in $F$ since they are inversely proportional.

\subsection*{R-charges}
Keeping this trivial construction aside, we go back to the full expression \eqref{Dn sigmas} now and try to figure out the non-canonical R-charges of the matter fields\footnote{The gravity duals for such theories are not well-known or well-studied and someone really needs to put some effort and figure this correspondence out. \cite{HW,KapD,ADey,FPufu} might be a good start.}. (For simplicity, we continue our analysis with $n_F$ set to zero.) Let's look at the generic inequality (instead of \eqref{inequalities} for $y$'s) in such a case\cite{JKPS}:
\equ{|y_a-y_b+\half(Δ_A-Δ_B)|\leq \half(Δ_A+Δ_B).
\label{ineqmod}
}
We see clearly that $Δ_B$ decides saturation points on the positive $x$-axis and $Δ_A$ on the negative, thus corroborating our observation for choice of $x^±_*$ in the algorithm section. The easiest way to see this is to choose a particular $y$ to be the `baseline' such that all bifurcations of $y$'s are in first and third quadrant (forget Figure \ref{fig:D4ys} in this regard). Furthermore, we do not have to solve the matrix model again to get the feel of where all these $Δ$'s can appear in the volume formula. We can just look at these inequalities and the list of saturation points in the case of canonical R-charges. Why? Because the expressions for saturation points are directly related to the $\bar{\s}$'s. This hand-waving quick analysis gives us the following identification
\equ{\bar{\s}^+_a=2Δ_{B_a}\bar{\s}_a\,;\qquad\bar{\s}^-_a=2Δ_{A_a}\bar{\s}_a.
\label{initident}
}
The index $a$ is misleading but the point is that there are as many pairs of bifundamental fields in a $\wh{D}$-quiver as there are `independent' $\bar{\s}$'s so this identification is fool-proof (modulo linear combinations because a priori there is no reason that a single $\bar{\s}$ should be associated to a particular $Δ$).
\fig{h!}{\includegraphics[width=3in]{Dn.pdf}
\put(-25,25){$Δ_3$}\put(-25,45){$Δ_4$}\put(-203,47){$Δ_1$}\put(-205,26){$Δ_2$}\put(-165,42){$Δ_5$}\put(-115,45){$\cdots$}\put(-67,42){$Δ_n$}
\caption{$\widehat D_n$ quiver diagram with the R-charges for bifundamental fields identified.}
\label{fig:Dn Rcharges}
}

Looking at the bifurcation patterns of $y$'s for various $\wh{D}$-quivers, we can (ignore thoughts of linear combinations to) propose an explicit form for the R-charges (see Figure \ref{fig:Dn Rcharges})
\equ{Δ_{A_a}=\half -\tfrac{t_ζ}{\bar{\s}_a}\,;\qquad Δ_{B_a}=\half +\tfrac{t_ζ}{\bar{\s}_a}
}
with the following identifications to be used in \eqref{initident} (suppressing $±$ on $\bar{\s}$'s and $A,B$ on $Δ$'s below)
\eqsn{\bar{\s}_0 &→ Δ_0=\half \\
\bar{\s}_1 &→ Δ_2 \\
\bar{\s}_2 &→ Δ_1 \\
\bar{\s}_3,⋯,\bar{\s}_{n-2}(=\bar{\s}_{n-1}) &→ Δ_5,⋯,Δ_n \\
\bar{\s}_n &→ Δ_3 \\
\bar{\s}_{n+1} &→ Δ_4\,.
}
So the main formula \eqref{Dn sigmas} can now be rewritten using \eqref{initident} to incorporate the non-canonical R-charges (and conjectured to hold in case of generic ones):
\eqs{\frac{\text{Vol}(Y^7)}{\text{Vol}(S^7)} &=\frac{1}{16}\(\sum_{a=0,1}^{n} \frac{\gamma_{a,a+1}}{Δ_{A_{a}}Δ_{A_{a+1}}\bar \sigma_a\, \bar \sigma_{a+1}} +\sum_{a=0,1}^{n} \frac{\gamma_{a,a+1}}{Δ_{B_{a}}Δ_{B_{a+1}}\bar \sigma_a\, \bar \sigma_{a+1}}\) \nn
&=\frac{1}{16}\sum_{a=0}^{n} \frac{\(Δ_{A_{a}}Δ_{A_{a+1}}+Δ_{B_{a}}Δ_{B_{a+1}}\)\gamma_{a,a+1}}{Δ_{A_{a}}Δ_{B_{a}}Δ_{A_{a+1}}Δ_{B_{a+1}}\bar \sigma_a\, \bar \sigma_{a+1}}\,·
\label{Dn sigmas deltas}
}

\section{Discussion}\label{discussion}
We found a non-trivial generalization of the free energy (or volume of the corresponding 7-dimensional Sasaki-Einstein manifolds expected from AdS/CFT correspondence) formula for $\wh{D}$-quivers deformed by complex FI parameters $ζ_a$, including the relevant cone (polygon) construction. Such a generalization depends only on the `sum' of $ζ$'s imaginary parts in the large $N$ limit and we related this particular quantity $(t_ζ)$ to non-canonical R-charges ($Δ_a$) of the bifundamental fields defining the quiver theories. Of course, a similar generalization can be expected to hold for $\wh{A}$-quivers (see Appendix \ref{Asol Appendix}) and $\wh{E}$-quivers that is (hopefully) as easy to compute.

As already hinted in Section \ref{review} and following \cite{JKPS}, one can study non-trivial superpotential deformations and RG flows to different IR fixed points using these generic volume formulas. It would be even more interesting if one could see similar behaviour in the gravity theory corresponding to these quiver theories or construct some new explicit examples on the AdS side of this AdS/CFT correspondence. We plan to pursue these questions in future work.

\section*{\centering Acknowledgements}
DJ thanks C. P. Herzog for encouraging correspondence at initial stages of this work. DJ also gratefully acknowledges P. M. Crichigno for insightful discussions on what has been written here and what could have been written.\\
This work is supported by MOST grant no. 104-2811-M-002-026 and CTS at NTU.

\appendix
\section[\texorpdfstring{Roots of $\widehat A_{m-1}$ and $\widehat D_n$ Lie Algebras}{Roots of Affine A and D Lie Algebras}]{Roots of $\bs{\widehat A_{m-1}}$ and $\bs{\widehat D_n}$ Lie Algebras}\label{A and D Appendix}
\fig{h!}{\begin{tabular}{ccc}
\includegraphics[width=2.5in]{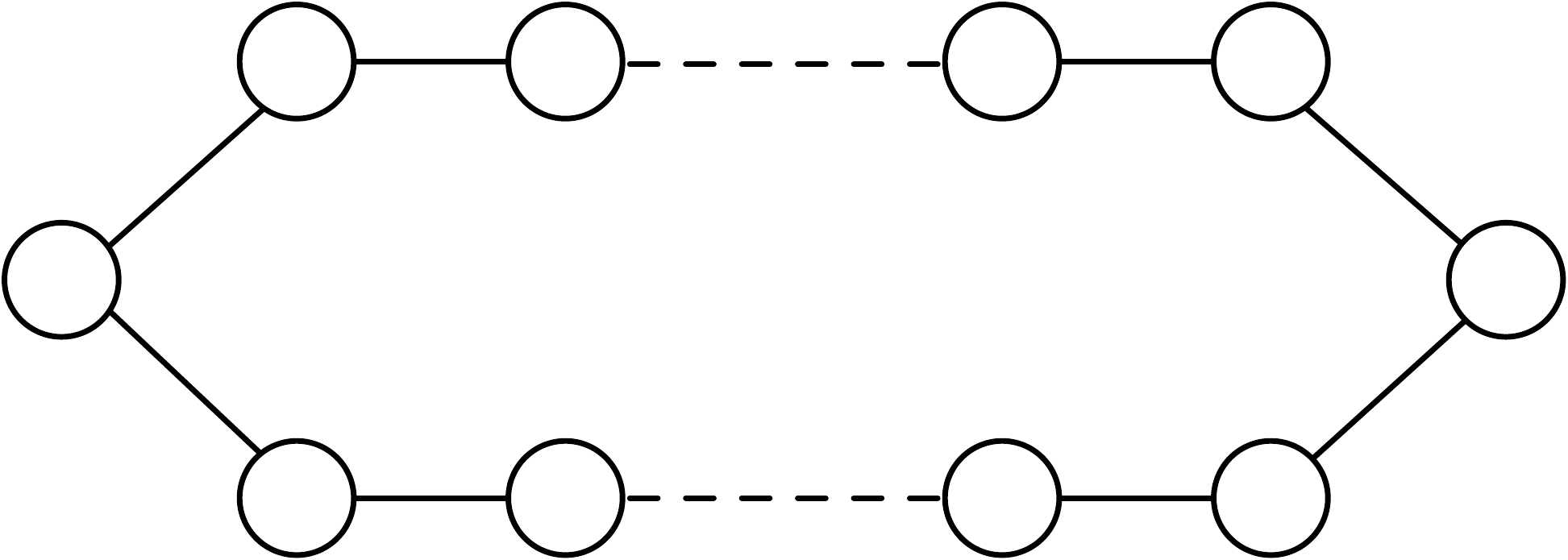} & \qquad\qquad &\includegraphics[width=2.5in]{Dn.pdf}
\put(-227,28){$\tilde{\alpha}_{j}$}\put(-420,28){$\tilde \theta$}\put(-380,-10){$\tilde{α}_{m-1}$}\put(-325,-10){$\cdots$}\put(-280,-9){$\tilde{α}_{k}$}\put(-380,67){$\tilde{α}_{1}$}\put(-325,67){$\cdots$}\put(-268,67){$\tilde{α}_{i}$}
\put(-10,-10){$\alpha_{n-1}$}\put(-12,67){$α_n$}\put(-180,67){$\theta$}\put(-180,-8){$α_1$}\put(-150,42){$α_{2}$}\put(-98,42){$\cdots$}\put(-52,42){$α_{n-2}$}
\end{tabular}
\caption{Dynkin diagrams for $\widehat A_{m-1}$ and $\widehat D_{n}$ Lie algebras.}
\label{fig:A and D quivers}
}
\noindent In Figure \ref{fig:A and D quivers} we show the affine Dynkin diagrams for the $\widehat A$ and $\widehat D$ Lie algebras along with the roots associated with every node. At each node, the CS level is given by $\tilde \alpha \cdot q$ and $\alpha \cdot p$ for $\widehat A$ and $\widehat D$, respectively. For $\widehat A_{m-1}$ we choose the following root basis
\equn{\tilde \alpha_a=e_a- e_{a+1}\,, \quad a=1,...,m-1\,; \qquad \tilde \theta=- e_1+ e_m,
}
where $e_a$  are canonical unit vectors of dimension $m$. For $\widehat D_n$ we choose
\equn{ \alpha_i =  e_i- e_{i+1}\,, \quad i=1,...,n-1\,; \qquad  \alpha_n= e_{n-1}+e_n\,, \qquad  \theta = -(e_1+e_2),
}
where $e_i$ are the unit vectors of dimension $n$.

\section[\texorpdfstring{A Construction for $\wh{A}$-quivers}{A Construction for Affine A-quivers}]{A Construction for $\bs{\wh{A}}$-quivers}\label{Asol Appendix}
We consider only odd $\wh{A}$-quivers here, \ie, when $m$ is even. (The case of odd $m$ is odd, to say the least.) Let us start with the definitions: $β_a= (1, q_a)$ for $a=1,⋯,m$, together with $\beta_0^±=\(0,n_F±2t_ζ\)$ and $\beta_{m+1}=\(0,1\)$.\footnote{In the usual treatment of pure $\wh{A}_{m-1}$, $β_{m+1}=-β_1$ but we find it difficult to introduce a $β_0$ consistently with that choice. We are not going to discuss here what the consequences of this different choice are on the nice identification of these $β$'s with the charges of $(p,q)$-branes.} Also, the constraint on $q$'s is solved by $q_m=-∑_{j=1}^{m-1}q_j$. Then the denominator factors $\sigma_a$ are written as follows
\equ{\sigma_a^{±}  = \sum_{b=0}^m \gamma_{a,b}\,; \qquad a=0,1,⋯,m+1.
\label{all A sigmas}
}
Realize that $\s_{m+1}^{±}=\s_{m+1}=m$. This leads to the compact volume formula for $\wh{A}$-quivers
\equ{\frac{\text{Vol}(Y^7)}{\text{Vol}(S^7)}=\frac{1}{4}\(\sum_{a=0}^{m} \frac{\gamma_{a,a+1}}{\sigma_a^-\, \sigma_{a+1}^-} +\sum_{a=0}^{m} \frac{\gamma_{a,a+1}}{\sigma_a^+\, \sigma_{a+1}^+}\)\!·
\label{An sigmas}
}
The cone construction is again straightforward and we do not spell it out here explicitly.

\fig{h!}{\includegraphics[width=2.5in]{An.pdf}
\put(-192,28){$k_1$}\put(-165,60){$k_2$}\put(-155,-8){$k_m$}\put(-98,60){$\cdots$}\put(-98,-2){$\cdots$}\put(2,28){$k_j$}
\put(-175,45){$Δ_1$}\put(-138,62){$Δ_2$}\put(-20,10){$Δ_j$}\put(-175,10){$Δ_m$}
\caption{$\widehat A_{m-1}$ quiver diagram with the CS levels and R-charges identified.}
\label{fig:An}
}
Proceeding as we did in the case of $\wh{D}$-quivers, the $\s$'s can be recast as (we again choose to set $n_F=0$ from here on)
\equ{\s^+_a=2Δ_{B_a}\s_a\,;\qquad \s^-_a=2Δ_{A_a}\s_a.
}
Again, staring at the bifurcation patterns of $y$'s for various $\wh{A}$-quivers, we propose the following explicit form for the R-charges (see Figure \ref{fig:An})
\equ{Δ_{A_a}=\half -\tfrac{t_ζ}{\s_a}\,;\qquad Δ_{B_a}=\half +\tfrac{t_ζ}{\s_a}
}
with the following identifications
\eqsn{\s_{0,m+1} &→ Δ_{0,m+1}=\half \\
\s_1,⋯,\s_{\sfrac{m}{2}-1} &→ Δ_1,⋯,Δ_{\sfrac{m}{2}-1} \\
\s_{\sfrac{m}{2}}=\s_{\sfrac{m}{2}+1} &→ Δ_{\sfrac{m}{2}+1} \\
\s_{\sfrac{m}{2}+2},⋯,\s_{m-1} &→ Δ_{\sfrac{m}{2}+2},⋯,Δ_{m-1} \\
\s_{m} &→ Δ_{\sfrac{m}{2}}\,.
}

We realize that we are missing $Δ_m$ but it is not the shortcoming of our approach, rather the structure of volume formula for $\wh{A}$-quivers betrays us. While for $\wh{D}$-quivers, the number of independent $\bar{\s}$'s is equal to the number of pairs of bifundamental fields (both being $n$), that number is not equal in the case of $\wh{A}$-quivers, which has $m$ pairs of such fields but only $m-1$ independent $\s$'s.\footnote{This ambiguity $(m≠m-1)$ persists even in the alternate choice of $β_{m+1}=-β_1$ because then $\s_{m+1}=\s_1$.} However, we know that in the case of ABJM $\big(\wh{A}_1\big)$, $Δ_2=Δ_1$ (as already mentioned in the main text) is required to obtain a match with the formula for generic R-charges derived in \cite{JKPS}. This has to do with the linear combinations of $Δ$'s which can appear in the inequality \eqref{ineqmod} since the first $y$ to be made redundant by a saturated inequality is still present in one more inequality (for example, $y_1$ appears in both $y_{12}$ and $y_{m1}$ inequalities). Since this issue did not arise in the case of $\wh{D}$-quivers (that's how $y$'s bifurcated), we did not worry about it there and we choose to not worry about it here too by just identifying $Δ_m=Δ_1$ (rest of the $y$'s saturate `nicely' so we have no more worries). Thus, we can finally rewrite the earlier formula \eqref{An sigmas} in terms of the non-canonical R-charges as follows
\equ{\frac{\text{Vol}(Y^7)}{\text{Vol}(S^7)} =\frac{1}{16}\sum_{a=0}^{m} \frac{\(Δ_{A_{a}}Δ_{A_{a+1}}+Δ_{B_{a}}Δ_{B_{a+1}}\)\gamma_{a,a+1}}{Δ_{A_{a}}Δ_{B_{a}}Δ_{A_{a+1}}Δ_{B_{a+1}}\sigma_a\,\sigma_{a+1}}\,·
\label{An sigmas deltas}
}
Since a lot of work has been done on $\wh{A}$-quivers, we have some concrete examples to compare the above relation with and we present two simple checks. First a trivial one: in ABJM theory, the above formula reduces to $\frac{1}{Δ_A^2Δ_B^2k}$ as expected from eq. (5.6) of \cite{JKPS} (we do need to use the fact that $Δ$'s are of the form $\half ± x$). Another less trivial check is with the coefficient $C$ in eq. (1.11) of \cite{Nosaka:2015iiw}. They pick a particular set of CS levels (we need to choose their parameters $p$ and $q$ to be equal because we have chosen to enforce $∑_jq_j=0$) which translate in our notation to: $q_1=⋯=q_{\sfrac{m}{2}}=\frac{k}{2}$ and the rest half being $-\frac{k}{2}·$ Their other two parameters become $ξ=-η=-\frac{2t_ζ}{k}$ (because $Δ_a=\half ± \frac{t_ζ}{pk}$). Using this dictionary, we find that \eqref{An sigmas deltas} reproduces $C$ correctly up to a redundant overall factor of $\frac{π^2}{2}·$

\references{
\bibitem{Moriyama}
S. Moriyama and T. Nosaka,
``Superconformal Chern-Simons Partition Functions of Affine D-type Quiver from Fermi Gas'',
\jhep{1509}{2015}{054} [\arXivid{1504.07710}{hep-th}].

\bibitem{Drukker}
B. Assel, N. Drukker and J. Felix,
``Partition Functions of 3d $\hat D$-quivers and their Mirror Duals from 1d Free Fermions'',
\jhep{1508}{2015}{071} [\arXivid{1504.07636}{hep-th}].

\bibitem{FGF1}
M. Mariño and P. Putrov,
``ABJM Theory as a Fermi Gas'',
{\it J. Stat. Mech.} {\bf 1203} (2012) P03001 [\arXivid{1110.4066}{hep-th}].

\bibitem{FGF2}
M. Mariño and P. Putrov,
``Interacting Fermions and N=2 Chern-Simons-Matter Theories'',
\jhep{1311}{2013}{199} [\arXivid{1206.6346}{hep-th}].

\bibitem{CHJ}
P. M. Crichigno, C. P. Herzog and D. Jain,
``Free Energy of $\widehat{D}_n$ Quiver Chern-Simons Theories'',
\jhep{1303}{2013}{039} [\arXivid{1211.1388}{hep-th}].

\bibitem{DMP}
N. Drukker, M. Marino and P. Putrov,
``From Weak to Strong Coupling in ABJM Theory'',
\cmp{306}{2011}{511} [\arXivid{1007.3837} {\color{cyan}\small [hep-th]}].

\bibitem{HKPT}
C. P. Herzog, I. R. Klebanov, S. S. Pufu and T. Tesileanu,
``Multi-Matrix Models and Tri-Sasaki Einstein Spaces'',
\pr{D}{83}{2011}{046001} [\arXivid{1011.5487}{hep-th}].

\bibitem{Anderson:2015ioa}
L. Anderson and J. G. Russo,
``ABJM Theory with Mass and FI Deformations and Quantum Phase Transitions'',
\jhep{1505}{2015}{064} [\arXivid{1502.06828}{hep-th}].

\bibitem{Hatsuda:2015gca}
Y. Hatsuda, S. Moriyama and K. Okuyama,
``Exact Instanton Expansion of the ABJM Partition Function'',
{\it PTEP} {\bf 2015} (2015) 11B104 [\arXivid{1507.01678}{hep-th}].

\bibitem{Russo:2015exa}
J. G. Russo and G. A. Silva,
``Exact Partition Function in $U(2) \times U(2)$ ABJM Theory Deformed by Mass and Fayet-Iliopoulos Terms'',
\jhep{1512}{2015}{092} [\arXivid{1510.02957}{hep-th}].

\bibitem{Nosaka:2015bhf}
T. Nosaka, K. Shimizu and S. Terashima,
``Large N Behavior of Mass Deformed ABJM Theory'',
2015, \arXivid{1512.00249}{hep-th}.

\bibitem{Nosaka:2015iiw}
T. Nosaka, ``Instanton Effects in ABJM Theory with General R-charge Assignments'',
2015, \arXivid{1512.02862}{hep-th}.

\bibitem{KWY}
A. Kapustin, B. Willett and I. Yaakov,
``Exact Results for Wilson Loops in Superconformal Chern-Simons Theories with Matter'',
\jhep{1003}{2010}{089} [\arXivid{0909.4559}{hep-th}].

\bibitem{HW}
A. Hanany and E. Witten,
``Type IIB Superstrings, BPS Monopoles, and Three-dimensional Gauge Dynamics'',
\npb{492}{1997}{152} [\arXivid{hep-th/9611230}].

\bibitem{KapD}
A. Kapustin, ``$D_n$ Quivers from Branes'',
\jhep{9812}{1998}{015} [\arXivid{hep-th/9806238}].

\bibitem{ADey}
A. Dey, ``On Three-dimensional Mirror Symmetry'',
\jhep{1204}{2012}{051} [\arXivid{1109.0407}{hep-th}].  
  
\bibitem{MaSp1}
D. Martelli and J. Sparks,
``The Large N Limit of Quiver Matrix Models and Sasaki-Einstein Manifolds'',
\pr{D}{84}{2011}{046008} [\arXivid{1102.5289}{hep-th}].

\bibitem{MaSp2}
D. Martelli and J. Sparks,
``AdS$_4$/CFT$_3$ Duals from M2-branes at Hypersurface Singularities and their Deformations'',
\jhep{0912}{2009}{017} [\arXivid{0909.2036}{hep-th}].

\bibitem{JKPS}
D. L. Jafferis, I. R. Klebanov, S. S. Pufu and B. R. Safdi,
``Towards the F-Theorem: N=2 Field Theories on the Three-Sphere'',
\jhep{1106}{2011}{102} [\arXivid{1103.1181}{hep-th}].

\bibitem{Gulotta2}
D. R. Gulotta, J. P. Ang and C. P. Herzog,
``Matrix Models for Supersymmetric Chern-Simons Theories with an ADE Classification'',
\jhep{1201}{2012}{132} [\arXivid{1111.1744}{hep-th}].

\bibitem{Gulotta1}
D. R. Gulotta, C. P. Herzog and S. S. Pufu,
``From Necklace Quivers to the F-theorem, Operator Counting, and T(U(N))'',
\jhep{1112}{2011}{077} [\arXivid{1105.2817}{hep-th}].

\bibitem{Zaffaroni}
M. Porrati and A. Zaffaroni,
``M-theory Origin of Mirror Symmetry in Three-dimensional Gauge Theories'',
\npb{490}{1997}{107} [\arXivid{hep-th/9611201}].

\bibitem{FPufu}
D. Z. Freedman and S. S. Pufu,
``The Holography of $F$-Maximization'',
\jhep{1403}{2014}{135} [\arXivid{1302.7310}{hep-th}]. 

}
\end{document}